\documentclass[twocolumn,times]{aastex63}
\usepackage{amsmath}
\usepackage{color}

\newcommand{\Chandra}{\textit{Chandra}}

\newcommand{\focc}{$f_{\text{occ}}$}
\newcommand{\factive}{$f_{\text{active}}$}

\newcommand{\lmxb}{$L_{\text{LMXB}}$}
\newcommand{\hmxb}{$L_{\text{HMXB}}$}

\begin{document}

\shorttitle{NUCLEAR X-RAYS IN LSBGs}
\shortauthors{HODGES-KLUCK ET AL.}

\title{Nuclear X-ray Activity in Low-Surface-Brightness Galaxies: Prospects for Constraining the Local Black Hole Occupation Fraction with a Chandra Successor Mission}

\author{Edmund J. Hodges-Kluck}
\affil{Code 662, NASA Goddard Space Flight Center, Greenbelt, MD 20771, USA}
\email{edmund.hodges-kluck@nasa.gov}

\author{Elena Gallo}
\affil{Department of Astronomy, University of Michigan, Ann Arbor, MI 48109, USA}

\author{Anil Seth}
\affil{Physics \& Astronomy, University of Utah, Salt Lake City, UT 84112, USA}

\author{Jenny Greene}
\affil{Princeton University, Department of Astrophysical Sciences, 4 Ivy Lane, Princeton University, Princeton, NJ 08544, USA}

\author{Vivienne Baldassare}
\affil{Yale University, Department of Astronomy, 52 Hillhouse Avenue, New Haven, CT 06511, USA}

\begin{abstract}
About half of nearby galaxies have a central surface brightness $\ge$1 magnitude below that of the sky. The overall properties of these low-surface-brightness galaxies (LSBGs) remain understudied, and in particular we know very little about their massive black hole population. This gap must be closed to determine the frequency of massive black holes at $z=0$ as well as to understand their role in regulating galaxy evolution. 
Here we investigate the incidence and intensity of nuclear, accretion-powered X-ray emission in a sample of 32 nearby LSBGs with the \textit{Chandra X-ray Observatory}. A nuclear X-ray source is detected in 4 galaxies (12.5\%). Based on an X-ray binary contamination assessment technique developed for normal galaxies, we conclude that the detected X-ray nuclei indicate low-level accretion from massive black holes. The active fraction is consistent with that expected from the stellar mass distribution of the LSBGs, but not their total baryonic mass, when using a scaling relation from an unbiased X-ray survey of normal galaxies. This suggests that their black holes co-evolved with their stellar population. In addition, the apparent agreement nearly doubles the number of galaxies available within $\sim$100~Mpc for which a measurement of nuclear activity can efficiently constrain the frequency of black holes as a function of stellar mass. We conclude by discussing the feasibility of measuring this occupation fraction to a few percent precision  below $\lesssim 10^{10} M_{\odot}$ with high-resolution, wide-field X-ray missions currently under consideration.  
\end{abstract}


\section{Introduction}
\label{section.intro}

A census of massive black holes (MBHs) in the nuclei of local galaxies is an important quantity for several reasons. First, it provides the present-day boundary condition (the ``fossil record'') on models for the formation and growth of MBHs \citep{volonteri12}, and on behavior during galaxy mergers. Second, to the extent that MBHs co-evolve with their host galaxies \citep{kormendy13}, it probes the importance of ``feedback'' in regulating galaxy growth. Third, the presence of an MBH is relevant to understanding stellar and gas dynamics in galactic nuclei even without feedback. Fourth, it is relevant to source rates from gravitational wave observatories and other probes of physics in strong gravity.

The local frequency of nuclear MBHs can be defined in terms of the occupation fraction (\focc) which is the fraction of galaxies with nuclear MBHs regardless of their activity. In practice, \focc\ cannot be reliably measured because of the limitations of different methods. Direct dynamical measurements (using stars or gas) are the gold standard, but existing samples are very biased relative to the galaxy population \citep{vandenbosch2015}. Meanwhile, the ``active'' fraction (\factive) provides only a lower limit to \focc, and can be defined in different ways (e.g., through optical line ratios, broad optical lines, X-ray activity, bolometric luminosity, etc.).

Despite their limitations, statistical analyses with these methods have led to the conclusion that \focc$\approx 1$ for large galaxies ($\log M_* \gtrsim 10$). On the other hand, most galaxies are smaller than this, and here \focc\ is poorly known. This is largely because their MBHs are less massive  \citep{gultekin09,kormendy13} making them hard to detect dynamically, although recent measurements suggest a high \focc \citep[but see][]{nguyen2018,nguyen2019}. Detecting accretion in these objects is challenging due to the presence of star formation and nuclear star clusters (NSCs), which are increasingly common in smaller galaxies \citep{seth08}. Using nuclear X-ray sources to trace MBHs, \citet{miller15} found that \focc$\ge$27\% over a mass range $8 < \log{M_*} < 11.5$, not ruling out 100\% even for small galaxies. Meanwhile, using spatially resolved optical spectroscopy to account for the contribution of starlight to the diagnostic line ratios, \citet{trump15} argued that \focc\ among low-mass galaxies is 10\% of that among the higher mass ones. The apparent inconsistency of these approaches indicates that more work is necessary to understand systematic effects and obtain a reliable \focc\ below $\log{M_*} \approx 10$. 

An additional, potentially complicating, factor is that many small galaxies have a surface brightness fainter than that of the night sky \citep[low surface brightness galaxies, or LSBGs;][]{impey97,vollmer13}. These galaxies may make up about half of nearby galaxies by number, but they are under-represented in catalogs and almost completely unexplored with regard to their MBH population.

LSBGs include galaxies of all types and with a large range of masses, but differ from their ``normal'' counterparts in a few ways. Notably, they tend to have very large gas fractions \citep[up to 95\%;][]{schombert01} and mass-to-light ratios, as well as low star-formation rates. LSBGs are also numerous, accounting for $\sim$50\% of nearby galaxies \citep{mcgaugh96,bothun97,dalcanton97,minchin04,haberzettl07}, and this makes them important for measurements of \focc. The formation of LSBGs remains an open and important question, but of particular importance here is that there appears to be no reason why they could not host MBHs at a similar rate as normal galaxies of the same dynamical mass, and their relatively slow evolution and lack of neighbors \citep{galaz11} may make them especially useful to distinguish between the ``light'' ($10^2-10^3 M_{\odot}$ Population~III remnants) and ``heavy'' ($10^4-10^6 M_{\odot}$ direct-collapse black holes) MBH seed hypotheses \citep{volonteri12}. There are few studies of MBHs in LSBGs, but there are hints that they tend to fall below the $M-\sigma$ relation, even in well developed bulges \citep{ramya11,subramanian16}. They are particularly under-studied in the X-rays; only a handful have been observed, and these were selected based on optical activity \citep{das09}. The majority of the work to identify AGNs in LSBGs has been done with optical line ratios \citep{schombert98,mei09,galaz11}.

Yet X-rays are important. High-resolution X-rays are sensitive probes of very low level accretion onto MBHs and relatively insensitive to dust absorption. The traditional cutoff for ``activity'' is at $L_{\text{bol}}/L_{\text{Edd}} > 10^{-3}$, with ``low luminosity'' AGNs at $L_{\text{bol}}/L_{\text{Edd}} > 10^{-5}$, but X-rays can probe down to $L_{\text{bol}}/L_{\text{Edd}} < 10^{-9}$ in local, massive systems. The main contaminating source of nuclear X-rays is from low- and high-mass X-ray binaries (XRBs), but a corrected \factive\ remains one of the best ways to search for nuclear MBHs. This formed the basis of the \textit{Chandra} X-ray Observatory \textbf{A}GN \textbf{MU}ltiwavelength \textbf{S}urvey of \textbf{E}arly-type galaxies programs \cite[AMUSE;][]{gallo08,miller12}, as well as several subsequent studies that expand to late-type galaxies \citep{foord17,she17,lee19}. One important result from these works is that there appears to be a simple relationship between $L_X$ and $M_*$ with some intrinsic scatter. The number of X-ray detected galaxies can then be compared to the number expected from this relation to constrain \focc\ \citep[a framework developed by][]{miller15}.

Thus, both to determine the X-ray nuclear properties of LSBGs, which have barely been studied, and to assess the potential to use them to constrain \focc\ and study MBH in an unbiased sample, we present a \Chandra{} survey of the nuclear activity in 32 LSBGs. The immediate scientific goal is to study the nuclear activity in LSBGs, as existing work is highly biased \citep[e.g.,][]{vandenbosch2015}, and it is timely to study their utility as future X-ray survey targets because of high-resolution X-ray concepts currently being studied. 

The remainder of this paper is organized as follows: Section~\ref{section.sample} describes the sample, Section~\ref{section.data} describes the observations and source detection method, and Section~\ref{section.xrbs} assesses the likelihood of contamination by X-ray binaries (XRBs). Section~\ref{section.results} presents the main result and discusses \factive\ in the context of other X-ray and LSBG studies. We argue that LSBGs are useful probes of \focc\ and present an observing strategy that includes them in Section~\ref{section.lynx}. We close by summarizing our findings in Section~\ref{section.summary}.

The distances adopted in this paper are based on the recessional velocity from the HyperLeda database \citep{makarov14} corrected for Virgo infall with a Hubble constant of 69.8~km~s$^{-1}$.

\begin{deluxetable*}{lllrrrrrrrr}
\tablenum{1}
\tabletypesize{\scriptsize}
\tablecaption{LSBG Sample Properties}
\tablewidth{0pt}
\tablehead{
\colhead{Name} & \colhead{Type} & \colhead{Active?} & \colhead{R.A.} & \colhead{Dec.} & \colhead{$d$} & \colhead{$\log M_{\text{HI}}$} & \colhead{$\mu_0(g)$} & \colhead{$M_g$} & \colhead{$(g-r)$} & \colhead{$\log M_*$} \\
& & & \colhead{(deg.)} & \colhead{(deg.)} & \colhead{(Mpc)} & \colhead{($M_{\odot}$)} & \colhead{(mag} & \colhead{(mag)} & \colhead{(mag)} & \colhead{($M_{\odot}$)}  \\
& & & & & & & \colhead{arcsec$^{-2}$)} & & & \\
}
\startdata
LSBC F570-04 	& Sa 	& N   & 168.23874 & 18.762 &  8.6 & ...   & 23.0$\pm$0.2 & -13.4$\pm$0.1	& 0.63$\pm$0.07	& 7.9$\pm$0.1 \\
LSBC F574-08 	& S0 	& N   & 188.15065 & 18.023 & 14.1 & ...   & 21.2$\pm$0.1 & -15.6$\pm$0.1	& 0.61$\pm$0.04	& 8.7$\pm$0.1 \\
LSBC F574-07 	& S0 	& ... & 189.87597 & 18.368 & 14.1 & ...   & 23.6$\pm$0.3 & -14.3$\pm$0.1	& 0.58$\pm$0.08	& 8.2$\pm$0.1 \\
LSBC F574-09 	& S0 	& N   & 190.5857  & 17.510 & 14.2 & ...   & 21.1$\pm$0.2 & -15.3$\pm$0.1	& 0.61$\pm$0.05	& 8.6$\pm$0.1 \\
IC 3605 	    & Sd/Irr& N   & 189.5873  & 19.541 & 14.3 & 8.55  & 22.0$\pm$0.1 & -15.1$\pm$0.1	& 0.19$\pm$0.06	& 7.9$\pm$0.1 \\
UGC 08839 	    & Im 	& H{\sc ii} & 208.85398 & 17.795 & 17.6 & 10.02 & 23.8$\pm$0.3 & -16.2$\pm$0.1	& 0.25$\pm$0.04	& 8.4$\pm$0.1 \\
UGC 05675 	    & Sm 	& N   &  157.12501 & 19.562 & 18.8 & 9.51  & 23.7$\pm$0.3 & -15.5$\pm$0.1	& 0.22$\pm$0.06	& 8.1$\pm$0.1 \\
UGC 05629 	    & Sm 	& N   & 156.05453 & 21.050 & 21.6 & 10.41 & 23.8$\pm$0.3 & -16.2$\pm$0.1	& 0.47$\pm$0.05	& 8.8$\pm$0.1 \\
LSBC F750-04 	& Sa 	& ... & 356.08417 & 10.118 & 23.8 & 8.39  & 23.0$\pm$0.2 & -15.5$\pm$0.1	& 0.39$\pm$0.08	& 8.3$\pm$0.1 \\
LSBC F570-06 	& S0 	& N   & 169.40918 & 17.818 & 24.8 & ...   & 22.3$\pm$0.1 & -16.8$\pm$0.1	& 0.67$\pm$0.04	& 9.3$\pm$0.1 \\
UGC 06151 	    & Sm 	& ... & 166.48456 & 19.826 & 24.8 & 8.79  & 22.2$\pm$0.2 & -17.2$\pm$0.1	& 0.41$\pm$0.04	& 9.0$\pm$0.1 \\
LSBC F544-01 	& Sb 	& ... & 30.33708  & 19.981 & 35.4 & 9.03  & 23.8$\pm$0.3 & -16.2$\pm$0.1	& 0.33$\pm$0.09	& 8.5$\pm$0.1 \\
LSBC F612-01 	& Sm 	& H{\sc ii} & 22.56423  & 14.678 & 36.8 & 9.00  & 23.7$\pm$0.3 & -16.0$\pm$0.1	& 0.34$\pm$0.09	& 8.5$\pm$0.1 \\
UGC 09024	    & S? 	& H{\sc ii} & 211.66891 & 22.070 & 38.8 & 9.35  & 20.8$\pm$0.1 & -18.1$\pm$0.1	& 0.32$\pm$0.04	& 9.3$\pm$0.1 \\
LSBC F743-01 	& Sd 	& ... & 319.68917 & 8.367  & 38.8 & 9.00  & 23.2$\pm$0.2 & -16.6$\pm$0.1	& 0.36$\pm$0.08	& 8.7$\pm$0.1 \\
LSBC F576-01 	& Sc 	& H{\sc ii} & 198.422   & 22.626 & 51.7 & 9.08  & 21.6$\pm$0.1 & -18.1$\pm$0.1	& 0.54$\pm$0.05	& 9.7$\pm$0.1 \\
LSBC F583-04 	& Sc 	& N   & 238.03887 & 18.798 & 57.4 & 8.90  & 23.9$\pm$0.3 & -17.5$\pm$0.1	& 0.46$\pm$0.07	& 9.2$\pm$0.1 \\
UGC 05005 	    & Im 	& H{\sc ii} & 141.12242 & 22.275 & 57.8 & 10.98 & 23.7$\pm$0.3 & -18.3$\pm$0.1	& 0.25$\pm$0.05	& 9.2$\pm$0.1 \\
UGC 1230 	    & Sm 	& H{\sc ii} & 26.38542  & 25.521 & 57.8 & 9.70  & 23.6$\pm$0.3 & -18.4$\pm$0.1	& 0.40$\pm$0.06	& 9.5$\pm$0.1 \\
UGC 04669 	    & Sm 	& H{\sc ii} & 133.77864 & 18.935 & 61.2 & 9.31  & 21.9$\pm$0.1 & -19.0$\pm$0.1	& 0.22$\pm$0.05	& 9.5$\pm$0.1 \\
UGC 05750 	    & SBd 	& H{\sc ii} & 158.93802 & 20.990 & 63.2 & 10.93 & 22.5$\pm$0.2 & -18.1$\pm$0.1	& 0.23$\pm$0.07	& 9.1$\pm$0.1 \\
UGC 4422 	    & SBc 	& AGN & 126.9251  & 21.479 & 64.6 & 9.91  & 19.8$\pm$0.1 & -21.4$\pm$0.1	& 0.57$\pm$0.01	& 11.0$\pm$0.1 \\
UGC 09927 	    & S0 	& AGN & 234.11572 & 22.500 & 67.9 & ...   & 19.11$\pm$0.04 & -19.9$\pm$0.1 & 0.81$\pm$0.03	& 10.7$\pm$0.1 \\
UGC 10017 	    & Im 	& N   & 236.39031 & 21.420 & 69.1 & 10.74 & 23.5$\pm$0.3 & -18.1$\pm$0.1	& 0.36$\pm$0.07	& 9.3$\pm$0.1 \\
UGC 10015 	    & Sd 	& H{\sc ii} & 236.41345 & 21.020 & 69.6 & 10.73 & 19.69$\pm$0.05 & -18.8$\pm$0.1	& 0.21$\pm$0.06	& 9.4$\pm$0.1  \\
UGC 3059 	    & Sd 	& AGN & 67.42687  & 3.682  & 69.6 & 10.00 & 22.4$\pm$0.2 & -21.2$\pm$0.1	& 0.23$\pm$0.05	& 10.3$\pm$0.1 \\
UGC 416 	    & Sd 	& H{\sc ii}   & 9.88753   & 3.933  & 70.2 & 9.93  & 22.4$\pm$0.2 & -18.5$\pm$0.1	& 0.45$\pm$0.06	& 9.6$\pm$0.1 \\
UGC 11578 	    & Sd 	& H{\sc ii} & 307.6785  & 9.190  & 70.6 & 9.98  & 22.3$\pm$0.2 & -19.2$\pm$0.1	& 0.33$\pm$0.04	& 9.7$\pm$0.1 \\
UGC 12845 	    & Sd 	& AGN & 358.9245  & 31.900 & 74.3 & 9.90  & 22.0$\pm$0.2 & -20.1$\pm$0.1	& 0.41$\pm$0.03	& 10.2$\pm$0.1 \\
UGC 11754 	    & Scd 	& H{\sc ii} & 322.38125 & 27.321 & 74.5 & 9.90  & 20.1$\pm$0.1 & -19.4$\pm$0.1	& 0.47$\pm$0.03	& 10.3$\pm$0.1 \\
LSBC F570-05	& S0 	& H{\sc ii} & 171.3237  & 17.808 & 74.5 & 9.61  & 20.5$\pm$0.1 & -19.5$\pm$0.1	& 0.67$\pm$0.04	& 10.4$\pm$0.1 \\
UGC 1455 	    & Sbc	& AGN & 29.7000   & 24.892 & 76.5 & 9.97  & 19.36$\pm$0.04 & -21.1$\pm$0.1 & 0.82$\pm$0.02 & 11.2$\pm$0.1 
\enddata
\tablecomments{\label{table.sample} LSBGs observed by \Chandra{} in this study. Activity is based on SDSS spectra or published claims of activity (see text), and nuclei with emission lines are classified as ``AGN'' or ``H {\sc ii}'' based on the \citet{Kewley2006} definition. Systems with no clear nuclear emission lines are marked ``N.'' Distances are from the HyperLeda database\citep{makarov14}, \ion{H}{1} masses are from \citet{huchtmeier89} and \citet{courtois09}, and stellar masses are computed from the SDSS $g$-band magnitudes and $g-r$ color (see text). Magnitudes reported here are in the AB system. We adopt a uniform uncertainty in the distance of 0.1~dex that propagates into the stellar mass. Some early-type galaxies have no \ion{H}{1} data.}
\end{deluxetable*}

\section{Sample}
\label{section.sample}

\subsection{Parent Sample}

We start with the \citet{schombert92} LSBG catalog, which was produced by searching the Palomar Sky Survey plates in the 3850-5500\AA\ band for galaxies fainter than the night sky. 
The advantage of using the \citet{schombert92} sample is that most of the galaxies have cataloged H~{\sc i} masses, which is important considering the tendency of LSBGs to have larger gas fractions than normal galaxies. However, the sample may be unrepresentative in a few ways. First, it does not include a strict cutoff in surface brightness and includes galaxies with ``normal'' central surface brightness but substantial, extended, LSB features. Second, the galaxies are almost all within $z<0.05$. \citet{rosenbaum09} found that LSBGs selected from the SDSS within this range tend to be dwarfs, whereas those at larger redshifts are luminous disks due to selection bias. Thus, we compared the \citet{schombert92} galaxies to more recent samples drawn from deeper exposures.

There is no single definition of an LSBG. The most common definition is an object whose central surface brightness $\mu_0 > 22$ or 23~mag~arcsec$^{-2}$ \citep{impey01}. For example, \citet{rosenbaum09} and \citet{galaz11} selected LSBGs with $\mu_0 > 22.5$~mag~arcsec$^{-2}$ from the Sloan Digital SKy Survey \citep[SDSS;][]{sdss12}. Other authors, such as \citet{greco2018}, define LSBGs based on their average surface brightness $\bar{mu}$, which includes nucleated galaxies with a ``normal'' $\mu_0$ but very low surface brightness disks \citep{bothun87,sprayberry95}. A variant on this approach is to define LSBGs based on the $\mu_0$ from a model profile after excluding the nuclear star cluster or active nucleus \citep[e.g.,][]{graham2003}.

Compared to these samples, the \citet{schombert92} galaxies are closer to Earth and tend toward the brighter end of the LSBG distribution, but are otherwise representative. Most, but not all, of these galaxies are regular dwarfs, and this is the population of most interest for \focc, and a key LSBG population to understanding the formation of LSB disks. It is also a good sample for an X-ray survey limited by the expected X-ray binary luminosity, considering that LSBGs are selected based on a broad observational, rather than physical, criterion.

\subsection{Working Sample}

We selected a subsample in order to compare \factive\ among LSBGs to normal galaxies in the AMUSE surveys. 
We adopted four criteria. First, we restricted the distance to $d<75$~Mpc to limit the exposure time required to achieve the same 0.3--10~keV $L_X \sim 10^{38}-10^{39}$~erg~s$^{-1}$ sensitivity as the AMUSE surveys. 159 galaxies in the \citet{schombert92} catalog meet this criterion, allowing for a 0.1~dex uncertainty in the distance. Secondly, we excluded galaxies without a well defined center in order to identify nuclear sources (about 35\% of systems). Thirdly, we excluded ``normal'' galaxies with minor LSB features included in the \citet{schombert92} catalog, but allowed nucleated and bulge-dominated galaxies with $\mu_0(g) < 22.5$~mag~arcsec$^{-2}$ as long as the average surface brightness within $D_{25}$ exceeded 23~mag~arcsec$^{-2}$.

Finally, we excluded galaxies with a total baryonic mass $\log (M_*+M_{\text{HI}}) < 7.5$ for consistency with the AMUSE survey. Here we use the total baryonic mass instead of $M_*$ because LSBGs tend to have high gas fractions whereas the gas fractions are very low for AMUSE galaxies, which are all early-type galaxies. The basis for the AMUSE restriction was concern that high-mass XRB (HMXB) contamination in late-type galaxies will be more severe than low-mass XRB (LMXB) contamination in early-type galaxies. However, LSBGs tend to have low SFR, and we show in Section~\ref{section.xrbs} that the potential for HMXB contamination is small. This also allows us to test whether the $L_X/M_*$ correlation found by \citet{miller12,miller15} applies to LSBGs, or whether the correlation is instead between $L_X$ and the total baryonic mass. However, as far as we know no galaxy was included that would not also meet a $\log M_* > 7.5$ threshold. After making these cuts, 83 galaxies remained.

To measure the surface brightness and the stellar mass we used $g$ and $r$ band optical data. We used the \ion{H}{1} masses from the \citet{huchtmeier89} and \citet{courtois09} catalogs. The main source of optical data was the SDSS \citep{sdss12}, but in multiple cases no SDSS data were available and we used the Pan-STARRS~1 DR2 \citep{Chambers2016}. We downloaded the calibrated galaxy images in $g$ and $r$ and fitted them with 2D S\'{e}rsic profiles using the {\tt Sersic2D} software from the astropy v4.0.1 Python library after masking surrounding point sources and obvious foreground or background objects coincident with the galaxy. The integrated $g$ band magnitudes and central surface brightness values are reported in Table~\ref{table.sample}.

About 30\% of systems from the \citet{schombert92} sample that meet our distance and identifiable center criteria have $\mu_0(g) < 22.5$~mag~arcsec$^{-2}$ for a single profile. Most are disky galaxies with a nuclear star cluster or other bright nuclear emission, and when allowing a second profile component for a nuclear point source the fits are improved and $\mu_0$ for the extended component typically falls into the LSBG threshold. However, some galaxies have a bright bulge surrounded by an extensive LSB disk or halo. In this case, adding a second profile component leads to one disky S\'{e}rsic component ($n<2$) and one spheroidal component ($n \sim 4$). We excluded galaxies where $\bar{\mu(g)}$ over $D_{25}$ is lower than 23~mag~arcsec$^{-2}$. Several galaxies in the remaining sample are also included in the \citet{graham2003} sample, who excised the central regions of nucleated sources to measure $\mu_0$, and our measurements are consistent with theirs.

We then used the integrated magnitude to estimate the stellar mass regardless of nuclear activity, following \citet{bell03} to calculate the mass-to-light ratio as $\log(M/L) = 1.519(g-r)-0.499$ for each $g$ band absolute magnitude. We adopt 5.11 as the absolute $g$ band magnitude of the Sun. The $g$ magnitudes, $g-r$ colors, and stellar masses of the galaxies are listed in Table~\ref{table.sample}. The statistical uncertainties in the measured magnitudes are small, so the uncertainty in $M_*$ comes primarily from uncertainty in the distances. We adopt a uniform 0.1~dex uncertainty for the distances throughout this paper, which are based on redshifts corrected for the Virgo infall. We do not include uncertainty from scatter in the $M/L$ relation.

Because it contains nearby, relatively bright LSBGs, the \citet{schombert92} catalog is already biased towards bright dwarf galaxies. The additional 75~Mpc distance cut does not materially change this. However, the criterion that each galaxy have a well defined center does bias the sample towards nucleated and spheroidal galaxies and against irregular galaxies. The mass cut also tends to exclude irregular galaxies and nearby dwarf ellipticals. On the other hand, and by design, this sample is well suited to the AMUSE-Field sample, which contains many normal dwarf galaxies with a similar mass range and is exclusively spheroidal.

After applying the mass cut at $\log M/M_{\odot} > 7.5$, a sample of 83 galaxies remained. We were awarded observing time on the \Chandra{} X-ray Observatory for 27 of these galaxies, which were selected based on the most efficient observing plan and \Chandra{} constraints. An additional five have existing \Chandra{} data. The \Chandra{} observation IDs and exposure times are summarized in Table~\ref{table.obs}.

\begin{figure}
\centering
\includegraphics[width=1.0\linewidth]{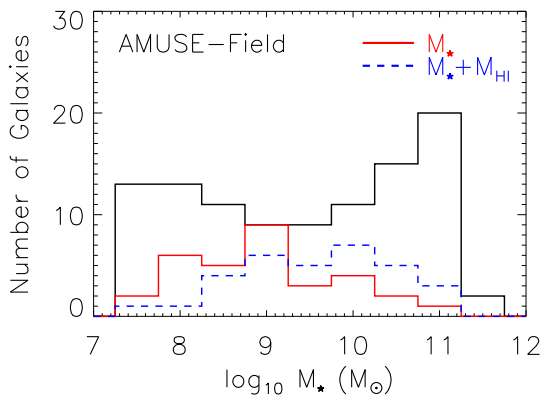}
\caption{The distribution of $M_*$ for the previously studied AMUSE-Field sample \citep[black histogram;][]{miller12} and the LSBG sample (red histogram), along with the distribution of $M_*+M_{\text{HI}}$ for the LSBG sample (blue dashed histogram). By design, the $M_*+M_{\text{HI}}$ sample is consistent with being drawn from the AMUSE-Field $M_*$ population (the latter includes only early-type galaxies with little gas).}
\label{figure.mstar_compare}
\end{figure}

The working sample includes 26 late-type galaxies and 6 early-type galaxies. 7/32 galaxies have $\log M_* > 10$, with the rest clustered around $\log M_* \sim 9.0$. The two-sided Kolmogorov-Smirnov (K-S) test indicates that the 33-galaxy sample has a mass distribution that is consistent with being drawn from the 83-galaxy sample ($p=0.26$). The K-S test also shows that the $M_* + M_{\text{HI}}$ distribution is consistent with being drawn from the AMUSE-Field $M_*$ distribution ($p=0.21$), but the $M_*$ distribution alone is not ($p=0.01$). Figure~\ref{figure.mstar_compare} shows these distributions. The gas fractions for most of the late-type galaxies are large, as expected for LSBGs.

The purpose of the AMUSE survey was to provide a view of nuclear activity unbiased by optical classification, but to compare our working sample to other LSBGs we investigated their optical nuclear properties. 22 of the 32 galaxies have SDSS spectra, of which 11 show clear emission lines that allow us to diagnose optical activity. Based on the pipeline line fluxes, only one (UGC~4422) has optical line ratios consistent with an AGN, but an additional four galaxies without SDSS spectra are candidate AGNs based on the \citet{schombert98} analysis, bringing the number of candidate AGNs to 5/22 (16\%). All of the AGN candidates are weak \citep{schombert98}, with $L < 10^{40}$~erg~s$^{-1}$. Meanwhile, 13/32 galaxies (41\%) have emission-line ratios consistent with star formation.  

To summarize, our working sample consists of 32 galaxies with X-ray observations. These galaxies tend to be nearby, brighter dwarf galaxies but also include some larger disk galaxies, and there are several AGN candidates. Compared to the larger LSBG population within $z<0.05$, these galaxies are more likely to be nucleated and tend to be more luminous than average \citep{greco2018}.  We return to the peculiarities of this sample when interpreting our results below.

\begin{figure*}
\centering
\includegraphics[width=1.0\linewidth]{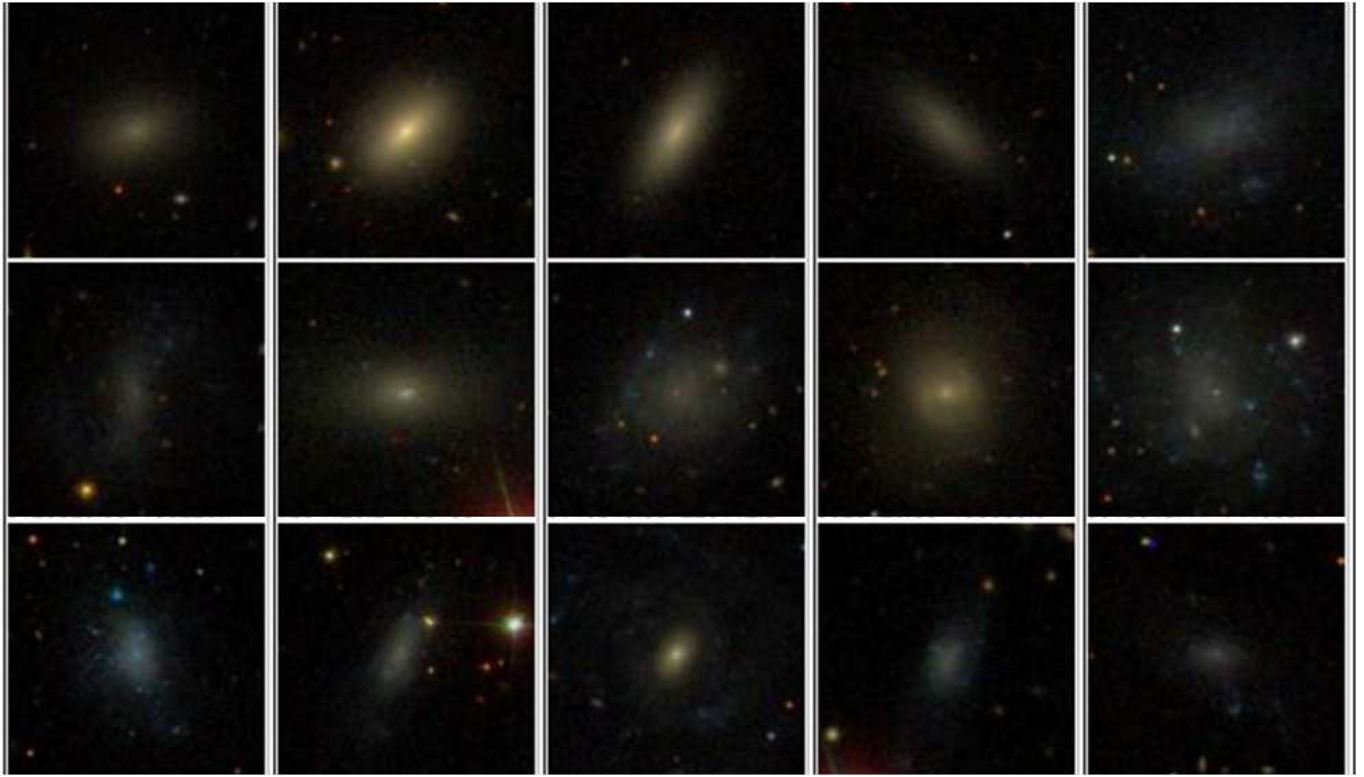}
\caption{The LSBGs in our sample span a range of masses, galaxy types, and morphology, as shown by these 15 LSBGs with SDSS snapshot images. 
}
\label{figure.sdss_gallery}
\end{figure*}

\section{Observations and Source Detection}
\label{section.data}

The new observations were obtained in the \Chandra{} Cycle 19 (2018) using the Advanced CCD Imaging Spectrometer (ACIS) camera. We centered each galaxy at the nominal aimpoint on the ACIS-S3 detector, which is back-illuminated and more sensitive to soft photons\footnote{http://cxc.harvard.edu/proposer/POG/}. The archival observations also used the ACIS-S3 detector. Observation information is listed in Table~\ref{table.obs}.

\begin{deluxetable}{llcr}
\tablenum{2}
\tabletypesize{\scriptsize}
\tablecaption{\Chandra\ Observation Log}
\tablewidth{0pt}
\tablehead{\colhead{Galaxy} & \colhead{ObsID} & \colhead{Date} & \colhead{$t_{\text{exp}}$ (ks)} \\}
\startdata
LSBC F570-04 	& 21006	& 2018-06-10	& 3.29\\
LSBC F574-08 	& 21008	& 2018-06-25	& 3.25\\
LSBC F574-07 	& 21009	& 2018-05-10	& 3.61\\
LSBC F574-09 	& 21012	& 2018-04-14	& 5.87\\
IC 3605 	    & 21016	& 2018-04-03	& 7.35\\
UGC 08839 	    & 21010	& 2018-04-03	& 3.44\\
UGC 05675 	    & 21011	& 2018-03-21	& 4.79\\
UGC 05629 	    & 21013	& 2018-07-02	& 6.07\\
LSBC F750-04 	& 21017	& 2018-08-26	& 8.93\\
LSBC F570-06 	& 21014	& 2018-11-25	& 6.75\\
UGC 06151 	    & 21015	& 2018-03-21	& 6.9\\
LSBC F544-01 	& 21019	& 2018-11-14	& 6.47\\
LSBC F612-01 	& 21020	& 2018-09-24	& 7.06\\
UGC 09024	    & 21018	& 2018-04-04	& 6.37\\
LSBC F743-01 	& 21021	& 2018-09-02	& 10.32\\
LSBC F576-01 	& 21022	& 2018-08-13	& 12.49\\
LSBC F583-04 	& 21023	& 2018-05-24	& 14.88\\
UGC 05005 	    & 21024	& 2018-06-19	& 5.77\\
UGC 1230 	    & 21025	& 2018-11-14	& 5.69\\
UGC 04669 	    & 21026	& 2018-05-23	& 6.66\\
UGC 05750 	    & 7766	& 2006-12-27	& 2.9\\
UGC 4422 	    & 21027	& 2018-03-21	& 6.71\\
UGC 09927 	    & 21028	& 2018-05-06	& 7.56\\
UGC 10017 	    & 21029	& 2018-05-17	& 7.86\\
UGC 10015 	    & 21030	& 2018-05-07	& 7.75\\
UGC 3059 	    & 7765	& 2007-01-01	& 3.3\\
UGC 416 	    & 21033	& 2018-09-09	& 11.21\\
UGC 11578 	    & 21031	& 2018-08-05	& 8.36\\
UGC 12845 	    & 7768	& 2007-02-18	& 3.25\\
UGC 11754 	    & 7767	& 2007-06-08	& 4.16\\
LSBC F570-05	& 21032	& 2018-06-28	& 9.53\\
UGC 1455 	    & 21032	& 2018-06-28	& 9.53
\enddata
\tablecomments{\label{table.obs} The \Chandra{} exposures are the sum of good-time intervals and are corrected for dead time. 
}
\end{deluxetable}

The data were processed and analyzed using the \Chandra{} Interactive Analysis of Observations (CIAO) v4.10 software\footnote{http://cxc.harvard.edu/ciao/}.We downloaded the primary and secondary data products and performed the standard recommended processing using the {\tt chandra\_repro} script, which filters out events with bad grades, identifies bad pixels, identifies good time intervals, and produces an analysis-ready level=2 events file. Most of the observations are very short, and none are significantly affected by particle background flares. 

The ACIS effective collecting area below 1~keV has degraded due to the build-up of molecular contamination on the filter window\footnote{http://cxc.harvard.edu/proposer/POG/}, and the decline has been particularly steep in the past few years. To optimize the sensitivity, the Cycle~7 data sets (obtained in 2007) were filtered to $0.3-8$~keV, while data sets from the past few years were filtered to $0.8-7$~keV. In Cycles 19 and 20, 90\% of the $0.3-8$~keV source X-ray events (counts) from a power-law spectrum with $\Gamma=1.5-2.5$ will fall in this bandpass (assuming no pileup and modest Galactic absorption), whereas only 50\% of the background will.

Source detection was performed using the CIAO Mexican-Hat wavelet {\tt wavdetect} script \citep{freeman02}. We used wavelet radii of 1, 2, 4, and 8 pixels, with an input map of the \Chandra{} psf for the ACIS-S3 chip constructed at $E=1.5$~keV for each observation. The other parameters were left as default. The source list was visually inspected to identify false detections (such as chip edges) and poorly separated sources. The filtered source list was then used with the CIAO {\tt wcs\_match} tool with the USNO-B1.0 catalog \citep{monet03} to align the images. In several cases, there were insufficient matches and we did not apply a correction. However, the typical correction is smaller than 1~arcsec, so we treat the astrometry as reliable for all exposures. 

Nuclear X-ray sources were identified as those sources for which the X-ray centroid error circle contains the position of the optical or IR nucleus, which also has some uncertainty. To estimate the uncertainty we used the centroid uncertainty from the best-fitting optical S\'{e}rsic profiles, which is generally a fraction of an arcsecond. This procedure finds three nuclear sources.

A second way to identify nuclear X-ray sources is to determine whether the number of counts in an $r=2$~arcsec aperture centered on the optical nucleus is higher than expected from the background. The half-power diameter of \Chandra{} with ACIS-S is about 0.8~arcsec, so events are concentrated within this region. However, roughly half of events are distributed between $r=0.4-2$~arcsec, so a true (but faint) source may not be identified by {\tt wavdetect}. With prior knowledge of where to look and a robust measurement of the background, such sources can be identified by comparison to the background rate. For most of the snapshot exposures, just three counts per aperture is sufficient to detect a source. The $0.8-7$~keV background rates expected in an $r=2$~arcsec aperture (based on a large region of blank sky) range from $1.5-2\times 10^{-5}$~counts~s$^{-1}$. The exposure times range from 3-11~ks, for which we expect an average of $0.05-0.2$~counts per aperture. Taking these as the averages in Poisson distributions, the odds of seeing three counts by random chance is less than 0.1\%. Since the nucleus positions are known, this is unaffected by the ``look elsewhere'' effect (although we note that other clusters of 3-4 counts detected with {\tt wavdetect} often do have catalog counterparts). However, an excess of counts does not necessarily imply a point source centered at the nucleus or a single point source. This procedure finds four nuclear sources, including the three found with {\tt wavdetect}. 

Three of the detected sources have 3-4 counts, including the one not found with {\tt wavdetect} (in UGC~9927). These are marginally detected in the sense that an integer number of counts must be detected and 2 counts is not significant. However, we estimated the likelihood of measuring 3 or more background counts in the nuclear apertures for our sample of 32 galaxies by simulating $10^8$ sets of observations with the average background in each aperture taken as the mean of a Poisson distribution. The odds of $N \ge 1$, 2, or 3 false positives are $P(N\ge 1) = 9\times 10^{-3}$, $P(N \ge 2) = 3\times 10^{-5}$, and $P(N \ge 3) = 2\times 10^{-7}$. 

The detected fraction depends on the energy bandpass, since the background is higher in the standard $0.3-8$~keV bandpass. In this case, neither source with 3 counts is significant. In addition, \Chandra{} ray-tracing simulations demonstrate that the concentration of events within the $r<2$~arcsec aperture is not a reliable way to distinguish sources and background, so apart from the small likelihood that the marginally detected sources are background fluctuations the spatial information is not useful. On balance, we conclude that the detected sources are astrophysical, and that at most one is a false positive. Additional observations would decisively settle the matter.

We converted the count rates and upper limits to $0.3-10$~keV luminosities by assuming a power-law spectrum with photon index $\Gamma=2$ and photoelectric absorption only from the Galaxy, using the Leiden-Argentine-Bonn survey\footnote{available at https://heasarc.gsfc.nasa.gov/cgi-bin/Tools/w3nh/w3nh.pl} \citep{kalberla05}. We ignore intrinsic absorption, but this will only lead to a small error as these are mostly face-on or early-type galaxies, for which we expect $N_{\text{H}} < 10^{21}$~cm$^{-2}$. At this column density, almost all absorption occurs below 0.8~keV where the ACIS-S effective area is very small. The number of counts in the detection cell and the $0.3-10$~keV luminosities or upper limits for each galaxy are given in Table~\ref{table.luminosities}.

This approach may not account for obscured AGNs. For example, sources with $N_{\text{H}} > 10^{23}$~cm$^2$ but $L_X < 10^{42}$~erg~s$^{-1}$ would not be detected. It is generally held that low luminosity AGNs (like the optical AGN candidates in our sample) lack such an obscuring torus, and anything as bright as $10^{42}$~erg~s$^{-1}$ would be a bright infrared source. Since none of the galaxies are included in the infrared AllWISE AGN catalog \citep{Secrest2015}, we have not missed any very obscured, luminous AGNs. On the other hand, high resolution infrared observations \citep[e.g.,][]{asmus2011} find some evidence for obscuring torii even in low luminosity systems, so we cannot rule this out. Such sources are unlikely to be found by increasing the X-ray sensitivity because their weak X-ray flux will be drowned out by the larger signal from X-ray binaries.

\begin{deluxetable*}{lrccccrrr}
\tablenum{3}
\tabletypesize{\scriptsize}
\tablecaption{LSBG Nuclear X-ray Properties}
\tablewidth{0pt}
\tablehead{
\colhead{Name} & \colhead{$\log M_*$} & \colhead{$f_{\text{nuc}}$} & \colhead{SFR$_{\text{nuc}}$} & \colhead{$\log L_{\text{LMXB}}$} & \colhead{$\log L_{\text{HMXB}}$} & \colhead{Counts} & \colhead{$\log L_X$} & \colhead{$\log P_{\text{XRB}}$} \\
 & \colhead{($M_{\odot}$)} &  & \colhead{($10^{-3} M_{\odot}$~yr$^{-1}$)} & \colhead{(erg s$^{-1}$)} & \colhead{(erg s$^{-1}$)} & & \colhead{(erg s$^{-1}$)} &
}
\startdata
LSBC F570-04 & 7.9  & 0.072$\pm$0.006 & 0.06$\pm$0.02 & 35.6$\pm$0.1 & 35.0$\pm$0.2 & 0   & $<37.6$ & $-2.68$\\
LSBC F574-08 & 8.7  & 0.113$\pm$0.004 & 0.32$\pm$0.05 & 36.7$\pm$0.1 & 35.6$\pm$0.1 & 0   & $<38.0$ & $-2.02$\\
LSBC F574-07 & 8.2  & 0.05$\pm$0.01   & 0.09$\pm$0.02 & 35.8$\pm$0.2 & 35.2$\pm$0.2 & 0   & $<38.0$ & $-2.86$\\
LSBC F574-09 & 8.6  & 0.10$\pm$0.03   & 0.8$\pm$0.2   & 36.3$\pm$0.1 & 36.1$\pm$0.2 & 0   & $<38.1$ & $-2.46$\\
IC 3605 	 & 7.9   & 0.06$\pm$0.02   & 1.35$\pm$0.09 & 35.7$\pm$0.2 & 36.3$\pm$0.1 & 0   & $<38.0$ & $-2.93$\\
UGC 08839	 & 8.4  & 0.014$\pm$0.004 & 0.9$\pm$0.2   & 35.6$\pm$0.1 & 36.2$\pm$0.2 & 0   & $<38.2$ & $-3.25$\\
UGC 05675 	 & 8.1  & 0.021$\pm$0.007 & 0.36$\pm$0.09 & 35.6$\pm$0.1 & 35.8$\pm$0.2 & 0   & $<38.4$ & $-3.52$\\
UGC 05629 	 & 8.8  & 0.023$\pm$0.007 & 0.4$\pm$0.1   & 36.1$\pm$0.1 & 35.6$\pm$0.2 & 0   & $<38.4$ & $-3.05$\\
LSBC F750-04 & 8.3  & 0.073$\pm$0.008 & 0.85$\pm$0.04 & 36.1$\pm$0.1 & 36.1$\pm$0.1 & 0   & $<38.3$ & $-2.88$\\
LSBC F570-06 & 9.3  & 0.056$\pm$0.004 & 0.42$\pm$0.04 & 37.0$\pm$0.1 & 35.8$\pm$0.1 & 2   & $<38.5$ & $-2.28$\\
UGC 06151 	 & 9.0  & 0.016$\pm$0.003 & 1.75$\pm$0.07 & 36.2$\pm$0.1 & 36.4$\pm$0.2 & 0   & $<38.5$ & $-3.05$\\
LSBC F544-01 & 8.5  & 0.06$\pm$0.01   & 7.2$\pm$0.6   & 36.2$\pm$0.1 & 37.0$\pm$0.2 & 0   & $<38.8$ & $-2.60$\\
LSBC F612-01 & 8.5  & 0.062$\pm$0.008 & 2.5$\pm$0.1   & 36.2$\pm$0.1 & 36.6$\pm$0.1 & 0   & $<38.8$ & $-3.02$\\
UGC 09024 	 & 9.3  & 0.077$\pm$0.003 & 14.4$\pm$0.2  & 37.1$\pm$0.1 & 37.3$\pm$0.1 & 0   & $<38.9$ & $-2.37$\\
LSBC F743-01 & 8.7  & 0.075$\pm$0.005 & 4.2$\pm$0.1   & 36.5$\pm$0.1 & 36.8$\pm$0.1 & 0   & $<38.7$ & $-2.64$\\
LSBC F576-01 & 9.7  & 0.117$\pm$0.004 & 11.1$\pm$0.3  & 37.6$\pm$0.1 & 37.1$\pm$0.1 & 0   & $<38.9$ & $-2.41$\\
LSBC F583-04 & 9.2  & 0.04$\pm$0.01   & 27$\pm$4      & 36.7$\pm$0.2 & 36.5$\pm$0.2 & 0   & $<38.9$ & $-3.13$\\
UGC 05005 	 & 9.2  & 0.020$\pm$0.006 & 4.4$\pm$0.1   & 36.6$\pm$0.1 & 36.8$\pm$0.1 & 0   & $<39.3$ & $-3.22$\\
UGC 1230 	 & 9.5  & 0.019$\pm$0.005 & 4.2$\pm$0.2   & 37.0$\pm$0.1 & 36.7$\pm$0.1 & 0   & $<39.3$ & $-3.33$\\
UGC 04669 	 & 9.5  & 0.038$\pm$0.007 & 11$\pm$2      & 37.4$\pm$0.1 & 37.1$\pm$0.2 & 4$^{+3}_{-1}$ & $39.6^{+0.2}_{-0.1}$  & $-3.18$\\
UGC 05750 	 & 9.1  & 0.06$\pm$0.01   & 21$\pm$3      & 36.9$\pm$0.2 & 37.5$\pm$0.1 & 1   & $<39.6$ & $-2.79$\\
UGC 4422\tablenotemark{a} 	 & 11.0 & 0.027$\pm$0.003 & 71$\pm$2      & 38.4$\pm$0.1 & 38.0$\pm$0.1 & 0   & $<39.3$ & $-1.94$\\
UGC 09927\tablenotemark{b} 	 & 10.7 & 0.05$\pm$0.02   & 12$\pm$3      & 38.3$\pm$0.2 & 37.2$\pm$0.2 & 3$^{+3}_{-1}$   & $39.5^{+0.3}_{-0.2}$  & $-3.09$ \\
UGC 10017 	 & 9.3  & 0.04$\pm$0.01   & 5.2$\pm$0.8   & 36.9$\pm$0.2 & 36.8$\pm$0.2 & 0   & $<39.3$ & $-3.24$\\
UGC 10015 	 & 9.4  & 0.040$\pm$0.007 & 12$\pm$3      & 37.9$\pm$0.1 & 37.2$\pm$0.2 & 1   & $<39.3$ & $-2.71$\\
UGC 3059\tablenotemark{a} 	 & 10.3 & 0.032$\pm$0.008 & 3.0$\pm$0.6   & 36.6$\pm$0.1 & 36.6$\pm$0.2 & 0   & $<39.6$ & $-3.77$\\
UGC 416 	 & 9.6  & 0.08$\pm$0.01   & 20.3$\pm$0.5  & 37.5$\pm$0.1 & 37.5$\pm$0.1 & 0   & $<39.2$ & $-2.44$\\
UGC 11578 	 & 9.7  & 0.038$\pm$0.008 & 11$\pm$2      & 37.6$\pm$0.1 & 37.2$\pm$0.1 & 3$^{+3}_{-1}$   & $39.5^{+0.3}_{-0.2}$  & $-2.96$\\
UGC 12845\tablenotemark{a} 	 & 10.2 & 0.024$\pm$0.005 & 11$\pm$1      & 37.9$\pm$0.1 & 37.2$\pm$0.2 & 0   & $<39.6$ & $-3.16$\\
UGC 11754\tablenotemark{a} 	 & 10.3  & 0.026$\pm$0.005 & 12$\pm$1      & 37.8$\pm$0.1 & 37.2$\pm$0.2 & 1   & $<39.5$ & $-3.13$\\
LSBC F570-05 & 10.4 & 0.053$\pm$0.003 & 36$\pm$8      & 38.5$\pm$0.1 & 37.7$\pm$0.2 & 1   & $<39.3$ & $-2.19$\\
UGC 1455\tablenotemark{b} 	 & 11.2 & 0.057$\pm$0.003 & 11$\pm$1      & 38.9$\pm$0.1 & 37.2$\pm$0.1 & 10$^{+4}_{-1}$  & $40.4\pm0.1$  & $-4.52$
\enddata
\tablenotetext{a}{The $0.3-8$~keV bandpass was used for detection.}
\tablenotetext{b}{Not found with {\tt wavdetect}.}
\tablecomments{\label{table.luminosities} 
The $M_*$ values are from Table~\ref{table.sample}, while $f_{\text{nuc}}$ refers to the fraction of $g$-band light in the nuclear aperture and SFR$_{\text{nuc}}$ is the nuclear SFR based on aperture-corrected GALEX photometry. $L_{\text{LMXB}}$ and $L_{\text{HMXB}}$ are the expected X-ray luminosities from low and high-mass X-ray binaries based on the nuclear stellar mass and SFR (see text). ``Counts'' refers to the number of X-ray counts detected in an $r=2$~arcsec aperture or with {\tt wavdetect} in the $0.8-7$~keV bandpass, and the luminosities have been converted to the $0.3-10$~keV bandpass assuming a power law spectrum with $\Gamma=2$. $P_{\text{XRB}}$ is the likelihood of detecting a total luminosity from LMXBs and HMXBs in the nucleus that exceeds $L_X$, accounting for the uncertainties and scatter in the X-ray luminosity functions.  
Errors on $f_{\text{nuc}}$ and the nuclear SFR are 1$\sigma$ statistical errors from photometry without including distance or scatter on the SFR indicator, whereas a 0.1~dex error was assumed for the distance and included in the error on $L_{\text{LMXB}}$ and $L_{\text{HMXB}}$. The uncertainty on the number of X-ray counts detected is based on \citet{Gehrels1986} and \citet{Ayres2004}. }
\end{deluxetable*}

\section{X-ray Binary Contamination}
\label{section.xrbs}

X-rays are excellent at identifying very low levels of nuclear MBH activity, but X-rays alone do not distinguish between weakly accreting MBHs and near-Eddington stellar-mass compact objects. A deep radio survey could do so, as stellar-mass objects are much more radio weak than MBHs \citep{merloni03}, but the necessary radio data do not yet exist. X-rays are also important counterparts, since there are radio contaminants as well (e.g., from star formation). Instead, we adopt a statistical approach based on \citet{foord17} and \cite{lee19} to assess the likely XRB contamination in the sample. 

XRB population studies in the Local Group and nearby galaxies have shown that the total luminosities of LMXBs and HMXBs in a galaxy correlate strongly with the stellar mass and star-formation rate (SFR), respectively\citep{gilfanov04,lehmer10,mineo12,lehmer16}. Since HMXBs cannot move far from star-forming regions in their lifetimes and LMXBs appear to be well distributed\citep[however, see][]{peacock16}, we can assume that the same correlations apply just to the nucleus. These correlations depend on the metallicity, which we take to be near-Solar. Then, from tracers of the stellar mass and SFR we can estimate the total nuclear \lmxb\ and \hmxb\ that could be confused with an accreting MBH. 

LMXBs and HMXBs are Poisson distributed and each follow an apparently universal X-ray luminosity function (XLF), which can be represented by a broken power law\citep{gilfanov04,mineo12}. Thus, the average XRB luminosities from the scaling relations can be converted into probability distributions from which we can determine the likelihood of detecting a total nuclear \lmxb\ or \hmxb\ at or above a given luminosity $P_{\text{XRB}}(L>L_0)$. In this case, $L_0$ could either be the observational sensitivity or the luminosity of a detected source. As the most likely non-XRB possibility is an accreting MBH, $P_{\text{MBH}} = 1 - P_{\text{XRB}}$ for any source. It is also useful to estimate the likelihood of detecting $N$ XRBs in the sample, which is calculated jointly from each $P_{\text{XRB}}(L>L_{\text{sens}})$ in the sample.

We implement this scheme using the \citet{lehmer10} expression for the 2-10~keV XRB luminosities:
\begin{eqnarray}
L_{\text{LMXB}} &=& (9.05\pm0.37)\times 10^{28} \text{ erg s}^{-1} \times M_* \\
L_{\text{HMXB}} &=& (1.62\pm0.22)\times 10^{39} \text{ erg s}^{-1}\times \text{SFR}
\end{eqnarray}
where $M_*$ and SFR are in units of $M_{\odot}$ and $M_{\odot}$~yr$^{-1}$, respectively. 
We adopt the \citet{gilfanov04} XLF for the LMXBs:
\begin{eqnarray}
dN/dL &= K_1 L^{-\alpha_1} & (L<10^{37}) \\
      &= K_2 L^{-\alpha_2} & (10^{37} < L < 10^{38.5}) \\
      &= K_3 L^{-\alpha_3} & (L>10^{38.5}) 
\end{eqnarray}
where $\alpha_1 = 1.0$, $\alpha_2 = 1.9$, and $\alpha_3 = 5.0$. The coefficients $K_1$, $K_2$, and $K_3$ are determined from $M_*$ such that \lmxb\ is consistent with the \citet{lehmer10} relation. The coefficients are slightly different in other studies \citep[e.g.,][]{gilfanov04}, but this has little impact on our results. The HMXBs follow a two-zone XLF in which $\alpha = -1.6$ between $10^{35}$ and $10^{40}$~erg~s$^{-1}$, and $\alpha \sim 3$ above $10^{40}$~erg~s$^{-1}$ \citep{mineo12}. The XLF slope changes somewhat when accounting for supersoft sources \citep{sazonov17}, but as we are insensitive to these sources the \citet{mineo12} values are sufficient.

We estimate the projected nuclear stellar mass from a nuclear aperture whose size is determined by the $r=2$~arcsec X-ray detection cell (or centroid error circle in the case of a detection). We include a small aperture correction and do not correct for any potential AGN, since at the low implied luminosities it is unclear whether most of the optical light comes from the AGN or a nuclear star cluster. The nuclear mass is estimated by calculating the fraction of light in this aperture and multiplying by the total stellar mass, assuming a uniform mass-to-light ratio. The nuclear mass fractions are given in Table~\ref{table.luminosities}.

We estimate the nuclear SFR from GALEX \citet{morrissey05} NUV ($\lambda$2300\AA) images in the same way using the \citet{kennicutt98} relation,  $\text{SFR} = 1.4\times 10^{-28} L_{\nu,\text{UV}}$~$M_{\odot}$~yr$^{-1}$, where $L_{\nu,\text{UV}}$ is in erg~s$^{-1}$~Hz$^{-1}$.
At 5.5~arcsec, the NUV PSF is considerably larger than the \Chandra{} (0.8~arcsec HPD) or SDSS (1.3~arcsec) PSF, so the aperture correction is more important. We correct for Galactic extinction using the  $E(B-V)$ value from NED, but not for unknown intrinsic extinction. The nuclear SFR values are listed in Table~\ref{table.luminosities}, where the uncertainty listed is statistical alone and assumes no scatter in the \citet{kennicutt98} relation and does not include uncertainty in the distance.

The nuclear $M_*$ and SFR, through the XRB scaling relations and XLF, yield the expected average number of nuclear XRBs per galaxy $\langle N_{\text{LMXB}} \rangle$ and $\langle N_{\text{HMXB}} \rangle$ (without mass matching). We then estimate the likelihood of detecting XRBs in a given galaxy by drawing $10^6$ Poisson deviates with $\langle N_{\text{LMXB}} \rangle$ and $\langle N_{\text{HMXB}} \rangle$ to simulate the range of possible numbers of XRBs. We randomly assign each XRB a luminosity by sampling the XLF, then sum the XRB luminosities to obtain a distribution of total nuclear $L_{\text{XRB}} = $\lmxb$+$\hmxb. We then calculate the likelihood of detecting nuclear X-rays from the XRBs, $P_{\text{XRB}}(L>L_X)$. Here $L_X$ refers either to the detected luminosity or the sensitivity in the event of a non-detection. 
These simulations take into account the uncertainty in the mass, SFR, and X-ray sensitivity or luminosity, which are dominated by uncertainty in the distance. We adopted a uniform 0.1~dex for this uncertainty. $P_{\text{XRB}}$ ranges from $10^{-4}$ to 0.02 for the galaxies in the sample (Table~\ref{table.luminosities}). The ranges for LMXBs or HMXBs alone are similar for the total sample, but differ from galaxy to galaxy.

The odds that $N\ge 1$ galaxies in our sample have detectable nuclear XRB emission are 0.071. The odds are 0.033 for LMXBs and 0.041 for HMXBs, individually. For HMXBs, any detectable emission is likely to be a single luminous ($L_X > 5 \times 10^{38}$~erg~s$^{-1}$) source, whereas for LMXBs a detection would imply multiple sources with $L_X \sim 10^{38}$~erg~s$^{-1}$, which would not necessarily appear point-like. A 7\% chance is not negligible, so we consider the impact of our assumptions.

We assumed that LMXBs follow the starlight rather than globular clusters. If not, then a nuclear star cluster may produce more LMXBs than expected from its luminosity and \lmxb\ would be underestimated. We have no way to assess this, but note that the X-ray detected fraction in nucleated galaxies is not particularly high \citep{foord17}. Secondly, we assume solar metallicity. \hmxb\ is higher for low metallicities, and we may have underestimated \hmxb\ by a factor of two \citep{douna15}. On the other hand, \lmxb\ is lower at low metallicities by a similar factor \citep{kim13}, so the net effect is minor for this sample. Thirdly, the FUV band is a better indicator of SFR, as early-type galaxies with almost no star formation can be bright in the NUV, which tends to overestimate \hmxb\. Unfortunately, FUV data are not available for all galaxies in the sample. However, the \citet{kennicutt98} relation is valid over a broad UV band, so this is likely a minor effect. Finally, the aperture correction for GALEX is large because its PSF is much larger than the nuclear aperture based on the \Chandra{} data. This increases the uncertainty in \hmxb.

Another potential issue is uncertainties in the XLF slopes. The normalizations are fairly well constrained \citep[e.g.,][]{lehmer16}, but there are signs that the XLF is not universal \citep{lehmer19}. For the \citet{gilfanov04} or \citet{mineo12} XLFs, most of the total luminosity is contained in the most luminous binaries. Since the odds of finding a luminous XRB in the nucleus are small, the more luminosity is contained in luminous sources the smaller the chance of contamination. Hence, steeper XLFs at the luminous end can actually increase $P_{\text{XRB}}$. There is not unlimited freedom here, since the XLF appears \textit{close} to universal. For the LMXBs, we adopted uncertainties of $\Delta \alpha_1 = 0.5$, $\Delta \alpha_2 = 0.2$, and $\Delta \alpha_3 = 1$ based on \citet{gilfanov04} and \citet{lehmer19}, whereas for the HMXBs we adopted uncertainties of $\Delta \alpha_1 = 0.25$ and $\Delta \alpha_2 = 0.5$ based on \citet{mineo12} and \citet{lehmer19}. We repeated the $P_{\text{XRB}}$ calculation by randomly (uniformly) varying the slopes within these envelopes over 1000 trials, which leads to a range of $0.01 < P_{\text{XRB}} < 0.13$ for the whole sample. Thus, it is likely that the uncertainties in distance, $M_*$, and SFR are more significant and also that all of the nuclear sources reported here are MBHs.

The most likely number of \textit{individually} detected XRBs above $L_{\text{sens}}$ in the whole sample, when considering entire galaxies (i.e., within $D_{25}$), is 3. The number of off-nuclear X-ray sources detected in this region in our sample is 4, which further supports the identification of the nuclear X-ray sources with MBHs. In Section~\ref{section.lynx} we discuss XRB contamination for higher sensitivity surveys. 

\section{Nuclear Activity in LSBGs}
\label{section.results}

A nuclear X-ray source is detected in 4/32 galaxies (\factive$=12.5$\%), or conservatively 3/32 (\factive$=9.4$\%) based on the discussion in Sections~\ref{section.data}. This active fraction is significantly lower than \factive\ reported in AMUSE-Virgo \citep[\factive$=68$\%;][]{gallo08}, AMUSE-Field \citep[\factive$=$45\%;][]{miller12}, or the Fornax cluster \citep[\factive$=$27\%;][]{lee19}. One possible reason is that the galaxies in our sample tend to have smaller $M_*$ (all of the detected sources in our sample occur in galaxies with $\log M* > 9$), which is supported by the \factive$=11.2$ measured in low-mass nucleated galaxies by \citet{foord17}. Since the total baryonic mass is consistent between the LSBG and AMUSE-Field samples, perhaps the relationship between $M_*$ and $L_X$ found by \citet{miller15} is indeed peculiar to stellar mass.

\begin{figure}
\centering
\includegraphics[width=1.0\linewidth]{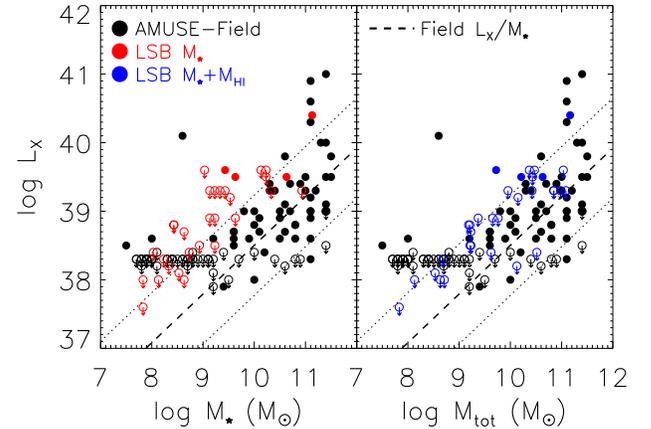}
\caption{\textit{Left}: $L_X$ plotted as a function of $M_*$ for the AMUSE-Field sample (black) and LSBG sample (red). Open circles are upper limits and filled circles are detected sources. The best-fit linear relation from \citet{miller12} is plotted as a dashed line, with the 1$\sigma$ scatter in dotted lines. \textit{Right}: The same as at left, except $L_X$ is plotted as a function of $M_* + M_{\text{HI}}$ for the LSBG sample (blue). The sensitivities for the LSBG sample were chosen based on $M_* + M_{\text{HI}}$.
}
\label{figure.lx_mstar}
\end{figure}

\begin{figure}
\centering
\includegraphics[width=1.0\linewidth]{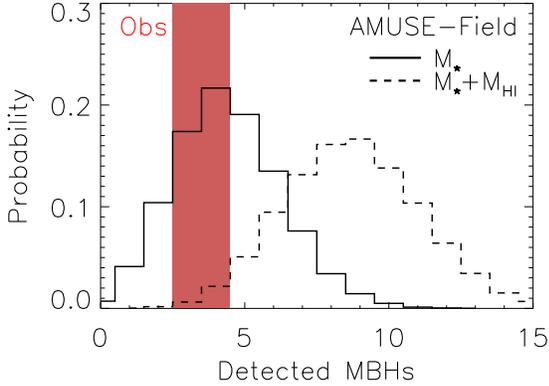}
\caption{The histograms show the expected number of X-ray detected MBHs for our sample, including the measured \Chandra{} sensitivities, if the $L_X - M_*$ relation from \citet{miller12} is correct (solid black line) and if the relation is instead $L_X - (M_*+M_{\text{HI}})$ (dashed black line). The observed number from the LSBG sample is shaded red. The width of each distribution is caused primarily by intrinsic scatter and secondarily by uncertainty in the slope, $M_*$, and $d$.}
\label{figure.field_prediction}
\end{figure}

There are too few LSBG sources to independently determine a relationship between $L_X$ and any galaxy property, but we can test this hypothesis by comparing the measured X-ray luminosities and upper limits in our sample to the AMUSE-Field sample, using either $M_*$ or $M_* + M_{\text{HI}}$. Figure~\ref{figure.lx_mstar} plots the detected LSBGs and upper limits on top of the AMUSE-Field results for both masses, and it is clear that there are too many undetected sources for the sensitivity if the LSBGs obey the best-fit AMUSE-Field relation,
\begin{equation}
\begin{split}
    \log L_X = & 38.4 - (0.04\pm0.12)+ \\
    & (0.71\pm0.10)\times(\log M_{\text{gal}} - 9.8) \\
    & \pm (0.73\pm0.09),
\end{split}
\end{equation}
where the last term is the intrinsic scatter, and $L_X$ depends on total baryonic mass. We can further use this relation to calculate the expected number of detected MBHs in the LSBG sample for either $M_{\text{gal}} = M_*$ or total baryonic mass. Figure~\ref{figure.field_prediction} shows the distributions of expected number of detected MBHs for a sample of the same size and with the same mass, distance, and sensitivity distribution as ours. The distributions account for the scatter in the AMUSE-Field relations and uncertainty in the masses and distances. Notably, if LSBGs follow the AMUSE-Field relation but the MBH luminosity is a function of total baryonic mass, there is only a 2.9\% chance of detecting four or fewer MBHs. On the other hand, there is a 22\% chance of detecting exactly four MBHs if the AMUSE-Field relation is instead particular to $M_*$. 

Of course, this does not prove that LSBGs follow the relation; a larger sample is needed to independently test this. Indeed, since all the detections occur in galaxies closer to $L_*$ than dwarfs, it is not clear whether the dwarf galaxies that make up a large proportion of nearby LSBGs differ from the more luminous galaxies that make up most of the more distant LSBGs. Nevertheless, if the AMUSE assumption that $L_X$ and mass are related in the same way at all masses is true, we can conclude that LSBGs follow this relation only if $L_X$ is related to the stellar mass rather than the total baryonic or dynamical mass.

We do not distinguish dependence on the total stellar mass or bulge mass. Prior studies of AGNs in LSBGs found that \factive\ increases with bulge luminosity \citep{mei09,galaz11}. The bulge contribution to the stellar mass in our sample varies strongly, but the four detected X-ray sources inhabit more massive galaxies whose bulges tend to be more massive relative to smaller galaxies (the one MBH candidate in an S0 galaxy, UGC~9927, is the least compelling detected source). A much larger X-ray study is needed to determine if \factive\ is correlated better with $M_*$ or $M_{*,\text{bulge}}$. 

Two of the four X-ray detected nuclei (UGC~9927 and UGC~1455) are categorized as AGN by \citet{schombert98}, albeit with low luminosities. This is consistent with the X-ray luminosities, all of which are below $10^{41}$~erg~s$^{-1}$. Three other galaxies in the sample (UGC~3059, UGC~4422, and UGC~12845) are also $L_*$ galaxies classified as AGN by \citet{schombert98} but are not detected in the X-rays.

Early studies of AGNs in giant spiral LSBGs found \factive$\sim$50\% \citep[e.g.,][]{schombert98}, but larger surveys including more galaxy types found a much lower \factive$\sim$5\% \citep{impey01,galaz11}. These surveys also find that LSBGs have lower \factive\ than normal (high surface-brightness) galaxies over a similar mass (or absolute magnitude) range, which \citet{galaz11} suggest is due to the low-density LSBG environments preventing the formation of bars or other instabilities that can fuel an AGN. These studies are based on optical emission-line diagnostics, which for our sample leads to \factive$\sim$15\% (5/32), which is likely because the sample is biased towards brighter dwarf galaxies \citep[especially compared to][]{galaz11} and includes some massive spirals. Our shallow X-ray survey finds \factive$\sim$10\%, and two of the four detected sources are in nuclei classified as star-forming. None of the detected systems would be classified as \textit{bona fide} X-ray AGNs.

Instead, the comparison with the AMUSE-Field sample indicates that weakly accreting MBHs in LSBGs are at least consistent with the high-surface-brightness galaxies of the same stellar mass. If LSBGs indeed show that there is a correlation between $L_X$ and stellar mass, but not baryonic or dynamical mass, this bears on black hole--galaxy co-evolution. In particular, we suggest that the inability of LSBGs to concentrate gas in the inner part of the galaxy is important to understanding their MBH growth. Although our sample is limited to relatively massive, isolated LSBGs, such a mechanism for limiting MBH growth would be relevant to most LSBGs.

\section{LSBGs and \focc}
\label{section.lynx}

Nuclear X-ray activity in LSBGs is consistent with that in normal galaxies of the same stellar mass, although a deeper, larger survey is needed to firmly establish the relationship between $L_X$ and $M_*$ in these systems. This makes LSBGs important to measuring \focc\ through the X-ray detection of weakly accreting MBHs, especially in the $\log M_*/M_{\odot} < 10$ regime where the heavy- and light-seed theories make different predictions. We emphasize that measuring \focc\ is valuable regardless of its ability to constrain formation theories (for which merger histories will also be important) because it is a probe of the total MBH mass density and anchors theories for co-evolution of MBHs with their host galaxies. 

In this section, we describe the logic behind an X-ray survey that could constrain the \focc\ to 1-5\% with a future wide-field, high resolution X-ray camera \citep[expanding on ideas explored in the Astro2020 Decadal Survey white paper by][]{gallo19}, or to $\sim$15\% with \Chandra{}. Then, we briefly explore how a survey could be constructed, including the expectation of many serendipitous LSBGs.

\subsection{Framework}

For a given $M_*$, \focc, and sensitivity the relation between the mean X-ray luminosity, $\bar{L}_X$, and $M_*$ predicts the measured \factive. For example, at a sensitivity 1$\sigma$ above $\bar{L}_X$, i.e., $\log L_{X,\text{thresh}} \log \bar{L}_X + 1\sigma$, one would expect \factive$=0.16$ at full occupation. Thus, measuring a lower-than-expected \factive\ would indicate \focc$<$1. In this case, one would need to detect zero sources in a sample of 26 galaxies to rule out \focc$=1$ at 99\% confidence. At a worse sensitivity of $\log L_{X,\text{thresh}} = \log \bar{L}_X + 2\sigma$, 200 galaxies are needed to draw the same conclusion. In general, the number depends on the cumulative distribution function. For galaxies covering a range in $M_*$ ($8 < \log M_* < 12$), one can simultaneously constrain $L_X/M_*$ slope(s), scatter, and the most likely \focc\ at each mass from the measured $L_X$ values and \factive. Using this approach, using 194 early type galaxies with \textit{Chandra} \cite{miller15} estimate \focc$>0.20$ below $M_*\gtrsim 10^{10} M_{\odot}$ (95\% credible interval). 

\begin{figure}
\centering
\includegraphics[width=1\linewidth]{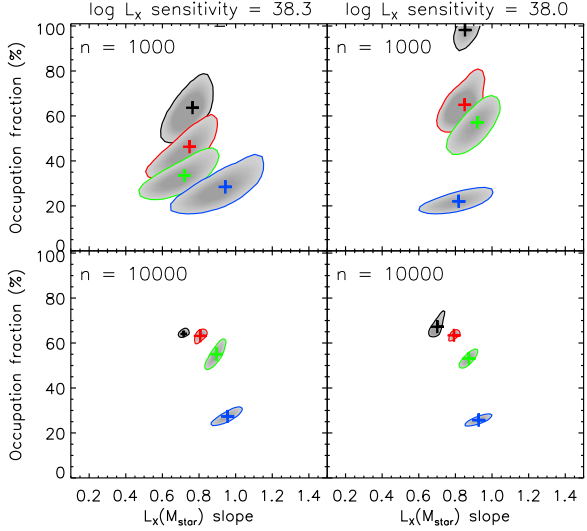}
\caption{Predictions for the constraints (posterior distributions) on \focc\ and $L_X/M_*$ slope for 1,000 and 10,000 galaxies (top and bottom rows), with sensitivity thresholds of $10^{38.3}$ and $10^{38}$~erg~s$^{-1}$, assuming a realistic mass distribution between $8 < \log M_*/M_{\odot} < 10$ in bins of 0.5~dex (blue, green, red, and black regions). The input $L_X/M_*$ and \focc\ in each mass bin for these simulations are from \citet{miller15}.}
\label{figure.focc_prediction}
\end{figure}

As a first step, we determined the number of galaxies needed to measure \focc\ to a precision of about 5\% assuming a power-law $L_X/M_*$ relation. We used a realistic mass distribution from \citet{blanton09} among bins 0.5~dex wide in $M_*$ from $8 < \log M_*/M_{\odot} < 10$, the $L_X/M_*$ slope of 0.8 from \citet{miller15}, and a uniform $L_{X,\text{thresh}}=1$ or $2\times 10^{38}$~erg~s$^{-1}$. The input \focc\ is a function of mass, ranging from 20\% at $\log M_*/M_{\odot} = 8$ to 100\% at $\log M_* = 10$, again following \citet{miller15}. Figure~\ref{figure.focc_prediction} shows the simulated posterior distributions for \focc\ and the $L_X/M_*$ slope for either 1,000 or 10,000 galaxies. With 10,000 galaxies, \focc\ is measured in these bins to 1-5\% precision. 

Fewer galaxies are needed when using a mass-dependent sensitivity (e.g., if $L_{X,\text{thresh}}-\bar{L}_X$ is constant). For a uniform $P_{\text{XRB}}$, the number of galaxies needed to overcome XRB contamination is proportional to $P_{\text{XRB}}$, since the inferred \factive\ will depend on $1-P_{\text{XRB}}$. So,
\begin{equation}
    N_{\text{gal}}^{\text{need}} \propto P_{\text{XRB}}\times \biggl[1-\text{CDF}(L_{X,\text{thresh}},\bar{L}_X)\biggr]
\end{equation}
where CDF is the normal cumulative distribution function for the case of Gaussian scatter. The feasibility of a tight \focc\ measurement depends on minimizing $N_{\text{gal}}^{\text{need}}$. As we show below, both high sensitivity and high angular resolution over a wide field are important. 

\subsection{Future X-ray Missions}

There are two mission concepts relevant to this work: \textit{Lynx} \citep{gaskin18} and the Advanced X-ray Imaging Satellite \citep[AXIS;][]{mushotzky18}. The \textit{Lynx} High Definition X-ray Imager (HDXI) has an effective collecting area of 20,000~cm$^2$ at 1~keV with a half-power diameter of $<1$~arcsec across the $23\times23$~arcmin field of view. AXIS is a similar instrument with 7,000~cm$^2$ effective area at 1~keV and $<$1~arcsec HPD across the $24\times24$~arcmin field of view. The high resolution is essential for two reasons. First, it enables the detection and centroiding of very weak, background-limited sources. Secondly, high resolution reduces confusion with individual luminous XRBs and reduces contamination by resolving out most of the luminosity. 

\subsection{XRB Contamination}

Whereas this study adopted a sensitivity threshold greater than $10^{38}$~erg~s$^{-1}$ to limit XRB contamination, similar snapshot exposures with the HDXI would achieve a sensitivity of $L_{X,\text{sens}} \sim 3\times 10^{36}$~erg~s$^{-1}$. This will lead to far more nuclear ``sources'' that are the sum of unresolved, lower luminosity XRBs, so we performed \textit{Lynx} and AXIS simulations to determine the impact, and how $P_{\text{XRB}}$ depends on distance $d$, resolution $\theta$, exposure time $t_{\text{exp}}$, and other factors. 

We simulated galaxies with $8 < \log M_*/M_{\odot} < 10$ in bins of 0.5~dex, with 10,000 galaxies per bin. We assumed that each galaxy is described by an exponential disk with a core radius $r_c = 2$~kpc that is independent of mass. We used the methods from Section~\ref{section.xrbs} to populate each galaxy with XRBs, which involves drawing a number of XRBs per galaxy and assigning positions and luminosities for each one. LMXB positions were randomly distributed weighted by the surface brightness, whereas HMXBs were randomly distributed within a 1~kpc radius for SFR ranging from $10^{-5}$ to 1~$M_{\odot}$~yr$^{-1}$ (i.e., star formation outside of 1~kpc of the nucleus was ignored as these HMXBs will not be a problem). The XRBs were randomly assigned luminosities weighted by the XLF.

We simulated HDXI and AXIS observations using the \textit{simx} software\footnote{https://hea-www.harvard.edu/simx/}, with the 2018 HDXI\footnote{http://hea-www.cfa.harvard.edu/~jzuhone/soxs/responses.html} and AXIS\footnote{http://axis.astro.umd.edu/} responses. We assumed an absorbed power law spectrum for each XRB, with $\Gamma=1.8$ and $N_{\text{H}} = 2\times 10^{20}$~cm$^{-2}$ (Galactic absorption). We then projected the galaxies to $d$ and selected $t_{\text{exp}}$ and $\theta$, assuming a circular Gaussian PSF where $\theta$ is the on-axis half-power diameter. The PSF distortion with off-axis angle can be described by a second Gaussian term. We consider the effect of PSF blurring below.

Sources are detected using {\tt wavdetect}, and for each XRB we compute the centroid error circle $\sigma = \sigma_{\text{telescope}}/\sqrt{N}$, where $N$ is the number of counts. We assume an optical galaxy centroid error of $\sigma=0.05$~arcsec, and reject any detected, non-nuclear XRBs. The accuracy of the centroid positions are insensitive to $\theta$. However, there is frequently a ``glow'' of X-rays from unresolved XRBs around the nucleus and from the wings of resolved XRBs. This glow is not uniformly distributed, but can be consistent with a weak nuclear point source and certainly impacts the centroid error circle. The proportion of galaxies with at least 5~counts within the nuclear aperture (using the 95\% encircled energy radius) from this glow is approximately linear in $\theta$. We compute $P_{\text{XRB}}$ by including the glow in the measured centroid error circle and counting galaxies as contaminated where there are at least 5~counts in the nuclear aperture from the glow. 

\begin{figure}
\centering
\includegraphics[width=1\linewidth]{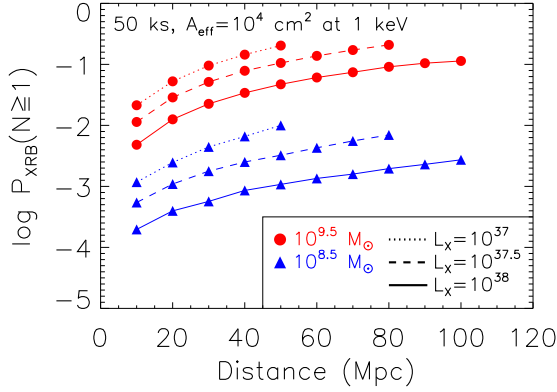}
\caption{The likelihood of detecting one or more nuclear XRBs (or enough counts from a diffuse ``glow'' to register as a detection) as a function of distance for $10^{9.5} M_{\odot}$ (red) and $10^{8.5} M_{\odot}$ (blue) galaxies and a \textit{Lynx} or AXIS-like mission. The different lines correspond to $L_{X,\text{thresh}}$ values of $10^{37}$ (dotted), $10^{37.5}$ (dashed), and $10^{38}$~erg~s$^{-1}$ (solid) and extend out to the distance to which such a source could be detected. See text for description of the simulation method.}
\label{figure.p_xrb}
\end{figure}

Figure~\ref{figure.p_xrb} shows the dependence of $P_{\text{XRB}}$ on $d$ for the cases of $\log M_*/M_{\odot} = 8.5$ and $9.5$, which represent the mass range of interest. This example uses an exposure time of 50~ks and the \textit{Lynx} spectral response (effective area as a function of energy), scaled to a collecting area of 1~m$^2$ at 1~keV. We computed $P_{\text{XRB}}$ over the range of parameters (assuming that SFR is proportional to mass, but not distributed in the same way) and find 
\begin{equation}
P_{\text{XRB}} \propto \theta \cdot t_{\text{exp}}^{-1/2} \cdot M_* \cdot d \cdot L_{X,\text{thresh}}^{-\beta}
\end{equation}
where $\beta \approx 1$ for the XLFs that we used. The dependence on $t_{\text{exp}}$ comes from resolving and rejecting more of the glow, while the dependence on $d$ is from the nuclear aperture covering a larger physical area in the galaxy. If $L_{X,\text{thresh}} \equiv L_{X,\text{sens}}$, then $P_{\text{XRB}} \propto t_{\text{exp}}^{\beta-1/2} d^{-1}$.

\subsection{Sensitivity and Number of Galaxies}

\begin{figure}
\centering
\includegraphics[width=1\linewidth]{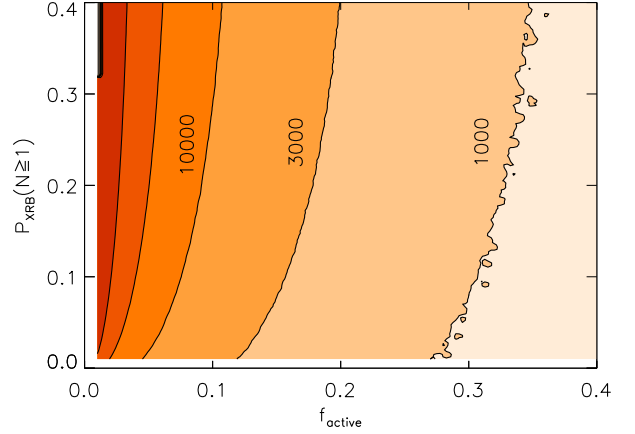}
\caption{The number of galaxies needed to measure a 68.3\% confidence interval equivalent to $\pm$5\% precision, as a function of \factive\ and $P_{\text{XRB}}$. To reduce $N_{\text{gal}}^{\text{need}}$ below about 1,000~galaxies in a given mass bin requires \factive$\gtrsim 0.3$, which will require higher sensitivity at lower $M_*$.}
\label{figure.ngal_focc}
\end{figure}

We may now determine the best $L_{X,\text{thresh}}$ at each $M_*$ to optimize $N_{\text{gal}}^{\text{need}}$. Figure~\ref{figure.ngal_focc} shows $N_{\text{gal}}^{\text{need}}$ as a function of \factive and $P_{\text{XRB}}$. Specifically, $N_{\text{gal}}^{\text{need}}$ is defined in Figure~\ref{figure.ngal_focc} based on achieving $\pm$5\% precision on \focc, for a 68.3\% confidence interval. At a given $M_*$, \factive\ must exceed 0.3 in order to keep $N_{\text{gal}}^{\text{need}}$ below 1,000. This approach can be generalized to measuring \focc\ in bins of $M_*$ 0.5~dex wide or to a continuous function \citep{miller15}. We use the former case to sketch the sensitivity requirements.

In a given mass bin, both \factive\ and $P_{\text{XRB}}$ increase with sensitivity. The increase is linear for $P_{\text{XRB}}$ but not for \factive; for \factive$\approx 0.25$, $(L_{X,\text{sens}}-\bar{L}_X)/\sigma \approx 0.5$, or well within the core of the distribution. Figure~\ref{figure.lx_mstar} shows that \Chandra{} achieved this for a sensitivity of $\sim 3\times 10^{38}$~erg~s$^{-1}$ at $\log M_*/M_{\odot} > 9.5$.  A reasonable approximation to the optimal sensitivity is $L_{X,\text{sens}} \approx \bar{L}_X$ (\factive$=0.5$ if \focc$=1$). This approximation is based on the fact that $N_{\text{gal}}^{\text{need}}$ decreases sharply as the sensitivity probes the core of the Gaussian distribution, but produce diminishing returns beyond. Meanwhile, $P_{\text{XRB}}$ is also a function of mass, so the $L_{X,\text{sens}} \approx \bar{L}_X$ applies at each mass bin. 

For three mass bins $\log M_*/M_{\odot} = 8.0-8.5$, $8.5-9.0$, and $9.0-9.5$, the AMUSE-Field $L_X/M_*$ relation predicts optimal sensitivities of $L_{X,\text{sens}} \approx 2$, 3, and $16\times 10^{37}$~erg~s$^{-1}$, respectively. At $\log M_*/M_{\odot} \le 9.5$, $P_{\text{XRB}} \le 0.1$ using the prior analysis (Figure~\ref{figure.p_xrb}). These considerations lead to a conservative estimate of $N_{\text{gal}}^{\text{need}} \sim 1000$ in each bin, or about 3000~galaxies overall at the low-mass end. 

We can infer the existence of sufficient targets within 100~Mpc, where short HDXI snapshots are sufficient. \citet{dobrycheva13} argue that there are about 37,000 galaxies in the SDSS in this volume, which covers 35\% of the sky. The luminosity function for a flux-limited sample, $\Phi(L) \propto (L/L*)^{-\alpha+3/2} e^{-L/L*}$ with $\alpha=-1.07$, implies that only about 10\% of the detected galaxies are at $8 < \log M/M_* < 9$ \citep{schechter76,binggeli88}. However, intrinsically there are more of these galaxies than the more massive ones, and the $\sim$10,000 detected in the SDSS in this range imply up to a factor of 3--10 more, depending on the slope of the luminosity function at $L \ll L*$ \citep[$\alpha < -1.3$;][]{blanton05,liu08}. 

Many of these will be LSBGs by definition, considering the SDSS sensitivity, which only make up 1.6\% of the SDSS spectroscopic sample \citep{galaz11}. In the next decade, the Vera Rubin Observatory \citep[VRO;][]{LSST09} will survey more than 20,000 square degrees down to $>27.5$~mag, so we expect at least 10,000 galaxies per mass bin. Although many will be unsuitable for observations (due to obscuration by the Galactic plane, proximity to bright sources, or morphology), there will easily be 1,000 candidate targets per bin. One challenge is that the photometric redshifts may not cleanly identify LSBGs within 100~Mpc \citep{greco2018}, so some spectroscopic follow-up will be necessary. 

\subsection{Strategy}

Observing 3,000 galaxies through pointed observations would require $\sim$100~Ms of HDXI time, or three years. Here we investigate the potential for serendipitous sources to reduce the dedicated observing burden to measure \focc. For the sake of argument, we assume two years of HDXI observations in a five-year mission with 75\% observing efficiency (with the rest of the time allocated to the \textit{Lynx} grating and microcalorimeter instruments). This amounts to 47~Ms. We further assume that the HDXI time is divided among \textit{long} (150~ks), \textit{medium} (50~ks), and \textit{short} (10~ks) exposures with no field overlap, with allocations of 20\%, 40\%, and 40\%, respectively. 

This would cover 10.5~deg$^2$, 63~deg$^2$, and 315~deg$^2$ for the long, medium, and short exposures. The sensitivities lead to distance limits, and thus to limiting volumes. At $8 < \log M_*/M_{\odot} < 8.5$, the limiting distances are 25~Mpc, 50~Mpc, and 100~Mpc for the short, medium, and long exposures. For $8.5 < \log M_*/M_{\odot} < 9.0$ they are 40~Mpc, 90~Mpc, and $>$100~Mpc, and for $\log M*/M_{\odot} > 9$ they are all $>$100~Mpc. Assuming that the fields are observed at random, a few hundred galaxies could be observed in the two higher-mass bins but only a few tens of galaxies in the low-mass bin. This is the most conservative estimate because it wrongly assumes a \textit{uniform} distribution, whereas a \Chandra{}-like observing plan will target denser regions.

\paragraph{Cluster Outskirts}

Galaxy clusters contain hundreds to thousands of galaxies and are of particular interest for X-ray observations. The cores of nearby clusters (Virgo, Fornax, Coma, and Perseus) have been well observed with \Chandra{}, largely to study the intracluster medium (ICM). Future observations of the Perseus or Coma cores will be less useful for measuring \focc\ because the ICM is so bright that reasonable exposures at HPD$=0.4$~arcsec will not be sensitive enough for galaxies with $\log M_*/M_{\odot} \lesssim 9.5$. In addition, \factive\ is lower in the Virgo core than in the field \citep{miller12}, which we expect to be an even stronger effect in the larger Perseus and Coma clusters. However, cluster outskirts remain under-studied and are a key area of interest for \textit{Lynx} and AXIS. AXIS is particularly interesting because of its planned low-Earth orbit \citep{mushotzky18}, which reduces the particle background and enables a cleaner study of accreting ICM at the outskirts. Tiled observations at the outskirts would likely capture a few thousand galaxies where the ICM surface brightness is low. These would be sensitive probes of \focc\ at $\log M_*/M_{\odot} \gtrsim 8.5$. LSBGs are an important part of this sample, as they make up a disproportionately large fraction of galaxies in clusters (likely due to ram-pressure stripping of gas). 

\paragraph{Deep Fields}

\citet{miller15} considered the role of deep fields; the 4~Ms \Chandra{} Deep Field-South probes AGNs in sub-$L_*$ galaxies in a cosmological volume, so assuming a uniform Eddington ratio distribution \citep{aird12}, they showed that the distribution of X-ray detections probes \focc. However, this is most effective above $\log M_*/M_{\odot} \ge 10$. \textit{Lynx} and AXIS would create fields of equivalent depth in exposures of a few hundred ks, which would result in tens of such fields in the first few years of either mission. The main benefit to measuring \focc\ at $\log M_*/M_{\odot} <10$ from the more distant objects is that the slope and scatter in the $L_X/M_*$ relation would be very tightly constrained, and possibly as a function of galaxy type or cosmological distance. 

\paragraph{Normal Galaxies}

Massive galaxies ($L>L^*$) are frequently targets of X-ray observations to study their hot gas, compact objects, or transient phenomena such as supernovae. However, dwarfs are clustered around more massive galaxies in the field \citep{binggeli90}, and based on their relative frequency we would expect each HDXI or AXIS field with a massive galaxy to have a number of dwarfs. Often, these will be unsuitable targets due to morphology or background, but especially within 100~Mpc galaxy observations will be important for building up a sample of $\log M_*/M_{\odot} < 9$ targets. \Chandra{} has observed numerous galaxies within this horizon, and we speculate that HDXI observations of these same galaxies would include at least 4,000 lower mass galaxies in fields with suitable sensitivity. 

It is worth noting that these observations would also allow the detection of X-rays from MBHs in stripped dwarf nuclei (frequently former nuclear star clusters), such as in the ultra-compact dwarf M60-UCD1 \citep{strader2013,seth2014}. A significant fraction of local MBHs (up to 1/3) may be located in such systems \citep{voggel2019}, and for relatively nearby galaxies they can be easily identified via VRO and Wide-Field Infrared Space Telescopes (WFIRST) colors \citep[using methods developed by][]{munoz2014}. We expect several around each galaxy relevant for the \focc\ measurement \citep[for the Milky Way, about 6 have been found;][]{kruijssen2019}, so a serendipitous sample of $\sim$1000 is easily feasible during the HDXI lifetime. 

\paragraph{Targeted Survey}

There will almost certainly be enough serendipitous sources at $\log M_*/M_{\odot} > 8.5$ to constrain \focc\ to 5\% precision, and so a major component of the program is ``free,'' requiring only that one waits several years. However, at the lowest masses it is much less certain that enough galaxies will be observed because the sensitivity of the typical field only captures systems within $d<25$~Mpc. There will not likely be enough deeper observations to make up for this limit by measuring \factive\ at a lower sensitivity. 

This motivates a snapshot survey of very nearby dwarf galaxies, many of which will be LSBGs. We estimate that 200-400 targets are required, with exposure times between 5-15~ks. This leads to a maximum total exposure time of $\sim$3~Ms. A dedicated survey of the Virgo cluster would significantly reduce the total time, since many of the nearby dwarf galaxies will be found in and around the cluster. If fields are selected to contain an average of two good candidates, the total observing burden is reduced to $\sim$1.5~Ms, which is a large program but a modest investment for measuring \focc. 

\section{Summary}
\label{section.summary}
We searched for nuclear X-ray sources in 32 nearby LSBGs with \Chandra\ and found 3-4, which we judge as very likely to be MBHs. This leads to \factive$=0.09-0.12$, which is consistent with the expectation from the best-fitting $L_X/M_*$ correlation from the AMUSE-Field study \citep{miller12}, which used \Chandra{} images of high surface brightness, early-type galaxies with almost no gas. However, \factive\ is inconsistent with the same relation if $M_*$ is replaced by the total baryonic mass, which is important since LSBGs have large gas fractions. 

This result suggests that weak nuclear activity innearby LSBGs with regular morphology is similar to that in normal galaxies of the same stellar mass, and thus that MBH growth is somehow tied to stellar, rather than baryonic or dynamical, mass. One explanation could be that isolated LSBGs are inefficient at concentrating gas that would lead both to star formation and MBH growth. However, the sample size is too small to independently measure any relationship between $L_X$ and $M_*$ (or total baryonic mass) in LSBGs, and a deeper, more extensive X-ray survey is needed to do this. Such a survey would also be able to answer whether the nuclear activity is better correlated with bulge luminosity, as argued by \citet{galaz11} for LSBGs, or total stellar mass. Nevertheless, our result supports a scenario in which MBHs co-evolve with the stellar component, rather than forming prior to it or in a way that correlates with halo mass. 

The agreement with the AMUSE-Field $L_X/M_*$ correlation also suggests that LSBGs can be used to constrain the local \focc\ of MBHs, albeit with too few detected sources to independently measure an $L_X/M_*$ relationship. LSBGs provide many relatively isolated targets with $\log M_* < 9$, where \focc\ predictions differ between heavy- and light-seed theories of MBH formation. A dedicated program, spaced over about five years, with a new, high resolution, wide-field X-ray camera such as \textit{Lynx} or AXIS would enable a measurement of \focc\ to a precision of several percent, thereby providing a strong local boundary condition on all MBH formation and evolution models, and extending studies of black-hole feedback to the low-mass end of the luminosity function. 

\acknowledgments
The authors thank the reviewer for a careful and thoughtful review that substantially improved this manuscript. 

Support for this work was provided by the National Aeronautics and Space Administration through Chandra Special Project SP8-19003X. 

This research has made use of the NASA/IPAC Extragalactic Database (NED) which is operated by the Jet Propulsion Laboratory, California Institute of Technology, under contract with the National Aeronautics and Space Administration. We acknowledge the usage of the HyperLeda database (http://leda.univ-lyon1.fr).

\bibliographystyle{aasjournal}

\end{document}